# Tracking User Attention in Collaborative Tagging Communities

Elizeu Santos-Neto
Electrical & Computer Engineering
University of British Columbia
2332 Mail Mall – KAIS 4075
+1.604.827.4270
elizeus@ece.ubc.ca

Matei Ripeanu
Electrical & Computer Engineering
University of British Columbia
2356 Mail Mall – KAIS 4033
+1.604.822.7281
matei@ece.ubc.ca

Adriana Iamnitchi
Computer Science and Engineering
University of South Florida
4202 E. Fowler Ave, Tampa, FL
+1.813.974.5357
anda@cse.usf.edu

## ABSTRACT
Collaborative tagging has recently attracted the attention of both industry and academia due to the popularity of content-sharing systems such as CiteULike, del.icio.us, and Flickr. These systems give users the opportunity to add data items and to attach their own metadata (i.e., tags) to stored data. The result is an effective content management tool for individual users. Recent studies, however, suggest that, as tagging communities grow, the added content and the metadata become harder to manage due to increased content diversity. Thus, mechanisms that cope with increase of diversity are fundamental to improve the scalability and usability of collaborative tagging systems.

This paper analyzes whether usage patterns can be harnessed to improve navigability in a growing knowledge space. To this end, it presents a characterization of two collaborative tagging communities that target the management of scientific literature: CiteULike and Bibsonomy. We explore three main directions: First, we analyze the tagging activity distribution across the user population. Second, we define new metrics for similarity in user interest and use these metrics to uncover the structure of the tagging communities we study. The properties of the structure we uncover suggest a clear segmentation of interests into a large number of individuals with unique preferences and a core set of users with interspersed interests. Finally, we offer preliminary results that suggest that the interest-based structure of the tagging community can be used to facilitate content retrieval and navigation as communities scale.

## Categories and Subject Descriptors
H.1.1 [**General**]: Systems and Information Theory - *Information Theory*. H.3.5 [**Information Storage and Retrieval**]: On-line Information Services - *Web-based services.*

## 1. INTRODUCTION
*Collaborative tagging systems* are online communities that allow users to assign terms from an uncontrolled vocabulary (i.e., tags) to items of interest. This simple tagging feature proves to be a powerful mechanism for personal knowledge management (e.g., in systems like CiteULike [3]) and content sharing (e.g., in communities such as Flickr [25]). Recently, collaborative tagging systems have attracted massive user communities: Novak et al. [16] report that in January 2006 *Flickr* congregated about one million users. Similarly, *del.icio.us* reached one million users in September 2006 [26].

Although collaborative tagging is attracting increasing attention from both industry and academia, there are few studies that assess the characteristics of communities of users who share and tag content. In particular, little research has been done on the potential benefits of tracking usage patterns in collaborative tagging communities. Moreover, recent investigations have shown that, as the user population grows, the efficiency of information retrieval based on user generated tags tends to decrease [2].

Mining usage patterns is an efficient method to improve the quality of service provided by information retrieval mechanisms in the web context. For example, usage patterns can be harnessed to improve 'browsing experience' via recommendation systems or to predict buying patterns and consequently increase revenue of e-commerce operations [8][9][10][17][18][19][20].

This work is motivated by the following conjecture: usage patterns can be harnessed to present relevant, contextualized information and deal with the reduced navigability generated by informational overload in large tagging communities.

We present encouraging preliminary steps to substantiate the above conjecture: We characterize two collaborative tagging systems: (*CiteULike* [3] and *Bibsonomy* [4]) as a first step towards a model to represent user interests based on tagging activity. After introducing related work (Section 2), we present a formal definition for tagging communities (Section 3) and the communities and the data sets this study explores (Section 4). We then characterize tagging activity distribution among users (Section 5) and we investigate the structure of user's shared interests (Section 6). Finally, we present preliminary results on the efficacy of using contextualized attention based on the structure of shared interests to improve the navigability in the system (Section 7). Section 8 summarizes our findings and outlines future research directions.

## 2. RELATED WORK
Two types of techniques, *implicit* and *explicit,* are traditionally used to elicit user preferences in the Web context [1][6][15]. *Explicit* techniques are based on direct input from a user with respect to her preferences and interests (e.g., page rating scales, item reviews, categories of interest). *Implicit* techniques infer a definition of user interests from her activity, e.g., using client-side or service-side mechanisms such as browser plug-ins, client extensions, and server-side logs to track usage patterns. Clearly, each technique has its own advantages and limitations in terms of accuracy, cost to the user, privacy control, or ability to adapt to changes on user interests trends.

In a tagging community context, the tags themselves can be interpreted as *explicit* metadata added by each user. Additionally, observed tagging activity including the volume and frequency with which items are added, the number of tagged items, or tag vocabulary size can be harnessed to extract *implicit* information.

Due to the youth of collaborative tagging systems, relatively little work has been done on tracking usage and exploring contextualized user attention in these communities. However, several studies present techniques and models for collecting and managing user attention metadata in the wider web context without exploring tagging features [1][6][15]. These techniques include post processing of usage logs, tracking user input (e.g. search terms) and eliciting explicit user preferences. Other investigations are concerned with methods to use contextualized attention to improve web search [1][15].

As a first step to modeling user attention in tagging communities, it is necessary to characterize collaborative tagging behavior. In this respect, Golder and Huberman [5] study user activity patterns regarding system utilization and tag usage in *del.icio.us* – a social bookmarking tool that allows users to share and tag URLs. First, they observe a low correlation between the number of items in each user's bookmark list and the number of tags used by each user. Next, they discuss the models that could explain this lack of correlation and suggest it is an effect of shared knowledge and imitation in associating tags. Finally, the authors suggest that the *urn model* proposed by Eggenberger & Polya [14] is an appropriate model to derive the evolution of tag usage frequency on a particular item.

The urn model can be formulated as follows: consider an urn that contains two colored balls. Iteratively, a ball is drawn at random from the urn and returned to the urn together with a new ball of the same color. If this process is repeated a number of times, the fraction of balls of a particular color stabilizes. The interesting aspect of this model is that if the process is restarted, this fraction converges to a different number. Golder and Huberman argue that this model captures the evolution of tag proportion observed in the *del.icio.us* data set. In studies related to contextualized user attention, this model may be valuable to predict future user tagging assignments which can be a useful input to recommendation mechanisms. Golder and Huberman's study, however, is limited in scale: their results on tagging behavior dynamics rely on only four days of tracked activity.

Other authors follow different approaches to investigate the characteristics of tagging systems. Schimtz [10][11] studies structural properties of *del.icio.us* and *Bibsonomy*, uses a tri-partite hypergraph representation, and adapts the small-world pattern definitions to this representation. Cattuto et al. [12] model usage behavior via unipartite projections from a tripartite graph. Our approach differs from these studies in terms of scale and in the use of dynamic metrics to define shared user interest: we define metrics that scale as the community grows and/or user activity increases (Section 6).

By analyzing *del.icio.us*, Chi and Mytkoswicz [2] find that the efficiency of social tagging decreases as the communities grow: that is, tags are becoming less and less descriptive and consequently it becomes harder to find a particular item using them. Simultaneously, it becomes harder to find tags that efficiently mark an item for future retrieval. These results indicate that, to facilitate browsing through tagging systems, it is increasingly important to take into account user attention in terms of observed tagging activity.

Niwa et al. [17] propose a recommendation system based on the affinity between users and tags, and on the explicit site preferences expressed by the user. Our study differs from this work as we use implicit user profiles and propose the use of entropy as a metric to characterize their effectiveness.

Outside the academic area, a number of projects explore the use of implicitly-gathered user information. We mention Google's initiative to explore users' past search history to refine the results provided by the Page Rank [8][9]. Commercial interest in contextualized user attention highlights that tracking user attention and characterizing collective online behavior is not only an intriguing research topic, but also a potentially attractive business opportunity.

## 3. BACKGROUND

A collaborative tagging community allows *users* to *tag items* via a web site. Users interact with the website by searching for items, adding new items to the community, or tagging existent items. The tagging action performed by a user is generally referred as a *tag assignment*.

For example, in *CiteULike* and *Bibsonomy*, each user has a library, i.e., a set of links to scientific publications and books. Each item in the library is associated with a set of terms (tags) assigned by users. It is important to highlight that, in both *CiteULike* and *Bibsonomy*, the process of assigning tags to items is collaborative, in the sense that all users can inspect other users' libraries and assigned tags. User can thus repeat tags used by others to mark a particular item. This is unlike other communities (e.g., *Flickr*) where each user has a fine-grained access control to define who has permissions to see the content and apply tags to it.

In *CiteULike* and *Bibsonomy* users have two options to add items to their libraries:

1. Browse the content of popular scientific literature portals (e.g. ACM Portal, IEEE Explorer, arXiv.org), to add publications to their own library, and

2. Search for items present in other users' libraries and add them to their own library.

While posting an item, a user can mark it with terms (i.e., tags) that can be used for future retrieval. The collaborative nature of tagging relies on the fact that users potentially share interests and use similar items and tags. Thus, while the tagging activity of one user may be self-centered the set of tags used may facilitate the job of other users in finding content of interest.

We represent a *collaborative tagging community* by the tuple: ***C=(U,I,T,A)***, where ***U*** represents the set of users, ***I*** is the set of items, ***T*** is the tagging vocabulary, and ***A*** the set of tag assignments.

The set of tag assignments is denoted by $A = \{(u, t, p) \mid u \in U, t \in T, p \in I\}$. From this definition of tag assignments, we can derive the definition of an individual user, item and tag, as follows:

- A user $u_k \in U$ is denoted by a pair $u_k = (I_k, T_k)$, where $I_k$ is the set of items user $k$ has ever tagged. Thus, an item $p \in I_k$ if and only if $\exists\ (u_k, t, p) \in A$, for any $t \in T_k$. Similarly, $T_k$ is the set of tags user $u_k$ applied before, where $t \in T_k$ if and only if $\exists\ (u_k, t, p) \in A$.

- An item $p_i \in I$ is denoted by $p_i = (U_i, T_i)$, where $U_i$ is the set of users who tagged this item, and $T_i$ is set of tags this item has received.

- A tag $t_j \in T$ is denoted by $t_j = (U_j, I_j)$, where $U_j$ is the set of users who used the tag $t_j$ before, and $I_k$ is the set of items annotated with the tag $t_j$.

## 4. DATA SETS AND DATA CLEANING

Both tagging communities we analyze: *CiteULike* [3] and *Bibsonomy* [4], aim to improve user's organization and management of research publications. Both provide functionality to import and export citation records in formats like BibTeX, for example.

The data sets analyzed in this article were provided by the administrators of the respective web sites. Thus, the data represents a global snapshot of each system within the period determined by the timestamps in the traces we have obtained (**Table 1**). It is important to point out that the Bibsonomy data set has timestamps starting at 1995, which we considered a bug. Moreover, Bibsonomy has two separate datasets, scientific literature and URL bookmarks. We concentrated our analysis on the scientific literature part of the data.

In the original *CiteULike* data set, the most popular tag is "*bibtex-import*" while the second most popular tag is "*no-tag*", automatically assigned when a user does not assign any tag to a new item. The popularity of these two tags indicates that a large part of users use *CiteULike* as a tool to convert their list of citations to BibTex format, and that users tend not to tag items at the time they post a new item to their individual libraries. Clearly, this is relevant information for system designers who might want to invest effort in improving the features of most interest.

Also, in *CiteULike* one user posted and tagged more than *3,000* items within approximately *5* minutes (according to the timestamps in the data set). Obviously, this behavior is due to an automatic mechanism.

**Table 1: Summary of cleaned data sets used in this study**

|  | CiteULike | Bibsonomy |
|---|---|---|
| **Period** | 11/2004—04/2006 | ??—12/2006 |
| **# Users (|U|)** | 5,954 | 656 |
| **# Items (|I|)** | 199,512 | 67,034 |
| **# Tags (|T|)** | 51,079 | 21,221 |
| **# Assignments (|A|)** | 451,980 | 257,261 |

Our objective is to concentrate only on those users who are using the system interactively to bookmark and share articles. Consequently, for the analysis that follows, we have the "robot" user (i.e., a user with 3,000 items tagged within 5 minutes) and users who used only the tags *bibtex-import* and/or *no-tag*. The total number of users removed from CiteULike represents approximately 14% of the original data set, while the users removed from Bibsonomy are around 0.6% of the original data set. **Table 1** summarizes the characteristics of each data set after the data cleaning operation.

## 5. TAGGING ACTIVITY

To gain an understanding on the usage patterns in these two communities, we start by evaluating the activity levels along several metrics: the number of items per user, number of tagging assignments performed, and number of tags used. The question answered in this section is the following:

*Q1: How is the tagging activity distributed among users?*

We aim to quantify the volume of user interaction with the system, either by adding new content to the community, or by tagging an existing item. Intuitively, one would expect that a few users are very active while the majority rarely interacts with the community.

Determining how often users perform tag assignments is important to help designing systems that track user attention. For example, in a context where activity information is used to recommend new items based on tag similarity, it can be necessary to compute the similarity level at the same rate as the rate with which new information is added into the system. Figure 1 presents the user rank according to the number of tag assignments performed during the time frame of our data set. In the results that follow, we present the data points observed together with a curve that provides a good model to the observed data (i.e. Hoerl function [21]). At the end of this section we comment more on the characteristics of this curve.

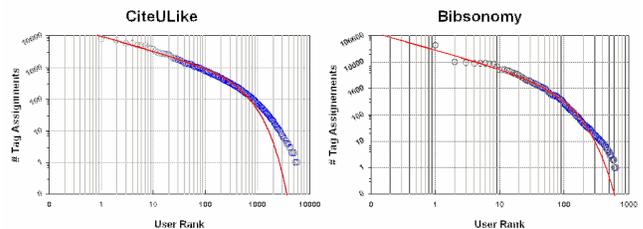

**Figure 1: User rank based on the number of tag assignments. Note the logarithmic scales on both axes.**

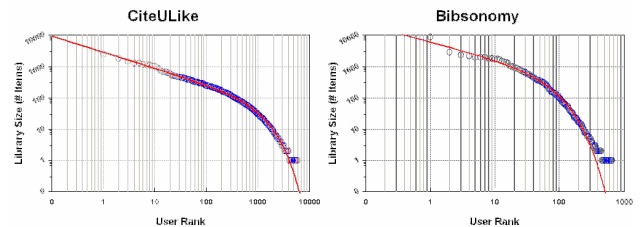

**Figure 2: User rank based on the library size.**

A second metric for tagging activity is the size of user libraries. Figure 2 plots user library size for users ranked in decreasing order according to the size of their libraries for CiteULike and Bibsonomy, respectively. This shows the size of the set of items a particular user pays attention to. The results confirm that the users in these two systems are heterogeneous in terms of activity intensity, as it has already been indicated by the tag assignment activity.

The correlation between a user's library size and her vocabulary is important to understand whether the diversity of the vocabulary used by each user grows with the number of items in her personal library. We observe that, in both communities the users' library and vocabulary sizes are strongly correlated for CiteULike ($R^2 = 0.98$, $n = 5954$) and less strongly, but still positively correlated, for Bibsonomy ($R^2 = 0.80$, n = 654). Although such correlation may seem intuitive, since users with a more diverse set of items would need more tags to describe them, this behavior is different from that observed by Golder and Huberman in del.icio.us [5]. A possible explanation is that in *del.icio.us* user is presented with tag suggestions based on past tagging activity when adding a new bookmark. These suggestions may bias and limit the size of a user vocabulary. However, further investigation is necessary to assess how a user vocabulary is affected by tagging recommendation.

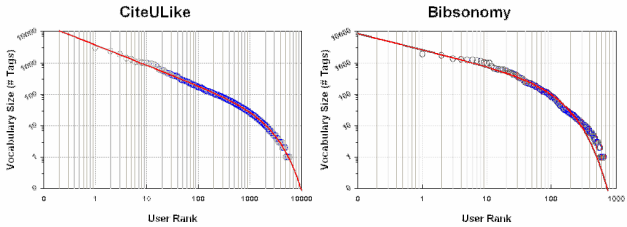

**Figure 3: User rank by vocabulary size**

A second finding is that the tagging activity (i.e., number of tagging assignments) and library size per user are strongly correlated for both communities (with $R^2$ above 0.97) while the correlations between the tagging activity and the vocabulary size is strong for CiteULike ($R^2 = 0.99$), but weaker for Bibsonomy ($R^2 = 0.67$).

A third finding is that tagging activity distributions are not well modeled by a Zipf-like distribution. Instead, a Hoerl model [21] that extends the power-law family and it is defined by Equation (1) fits better:

$$f(x) = ab^x x^c \qquad (1)$$

Table 2 contains the Hoerl parameters *a, b,* and *c* determined via a curve fitting process for each of the ranking distributions observed.

**Table 2: Coefficients determined for the Hoerl function**

| CiteULike | a | b | c |
|---|---|---|---|
| Tag Assignments | 9,767.13 | 0.9979 | -0.4754 |
| Library Size | 2,609.77 | 0.9988 | -0.4772 |
| Vocabulary Size | 3,338.55 | 0.9992 | -0.5964 |
| **Bibsonomy** | | | |
| Tag Assignments | 28,969.29 | 0.9864 | -0.6888 |
| Library Size | 6,137.49 | 0.9850 | -0.5461 |
| Vocabulary Size | 2,608.45 | 0.9907 | -0.5126 |

Similar to the Zipf distribution, the Hoerl function has been used to model a large number of natural phenomena. The most relevant to collaborative tagging is the use of Hoerl function to describe the distribution of bio-diversity across a geographic region [22][24]. Considering each user's library a *region* in a collaborative tagging community, one may draw a comparison between the potential *diversity* found in the users' library regarding the number of items in it, and the bio-diversity distribution across geographic regions.

Although a Hoerl function is a good fit for the activity distributions, this does not directly imply that diversity of user libraries or vocabularies represents a phenomenon which is similar to those presented by studies on biodiversity. Nevertheless, the Hoerl function does provide a good model for collaborative tagging activity and it can be useful to study user diversity in collaborative tagging systems in the future.

To summarize: in the communities we study, the intensity of user activity is distributed over multiple orders of magnitude, it is well modeled using the Hoerl function and, unlike in other communities, there is a strong correlation in activity in terms of items set and vocabulary sizes.

## 6. EVALUATING USER SIMILARITY

While the analysis above is important for an overall usage profile evaluation of each community, it provides little information about user interests. Assessing the commonality in user interests is important for identifying user groups that may form around content of common interest. Thus, a natural set of questions that we aim to answer in this section are:

Q2: *Is the tagging community segmented into several sub-communities with different interests? Do users cluster around particular items and tags?*

To address these questions, we define the *interest-sharing graph* after the intuition of data-sharing graphs introduced by Iamnitchi et al. [27]. An interest-sharing graph captures the commonality in user interest for an entire user population: Intuitively, users are connected in the interest-sharing graph if they focus on the same subset of items and/or speak similar language (i.e., share a subset of tags).

More formally, consider a graph $G = (U, E)$ where nodes are users and edges represent the existence of shared interests or activity similarity between users. The rest of this study explores three possible definitions for user interest or activity similarity. All these definitions employ a threshold $t$ for the percentage of items or tags shared between two users:

1) The *User-Item* similarity definition considers two users' interests similar if the ratio between the sizes of the intersection and the size of the union of their item libraries is larger than a threshold $t$. This is expressed by Equation 2.

$$e_{kj} \in E \Leftrightarrow \frac{|I_k \cap I_j|}{|I_k \cup I_j|} \qquad \textit{User-Item} \quad (2)$$

2) The *User-Tag* definition is similar to the definition above but considers the vocabularies of the two users rather than their libraries.

$$e_{kj} \in E \Leftrightarrow \frac{|T_k \cap T_j|}{|T_k \cup T_j|} \qquad \textit{User-Tag} \quad (3)$$

3) Unlike the *User-Item* definition in Equation 2 above, the *Directed User-Item* considers two users' interests similar if the ratio between the intersection of their item libraries and the size of one user library is larger than a threshold $t$. The idea is to explore the role played by users with large libraries via the introduction of direction to the edges in the graph.

$$e_{kj} \in E \Leftrightarrow \frac{|I_k \cap I_j|}{|I_k|} \qquad \textit{Directed-User-Item} \quad (4)$$

In our analysis of real tag assignment traces from the two tagging communities, even with low values for the sharing ratio threshold $t$, the final graph contains a large number of *isolated* nodes. Indeed, by setting the threshold as low as one single item (i.e., two users are connected if they share *at least* one item); we find that, in *CiteULike*, 2,672 users (44.87%) are not connected to any other user. This suggests that a large population of users has individual preferences.

Figure 4 presents, for the three similarity metrics defined above, the number of connected components for both CiteULike and Bibsonomy, for thresholds $t$ varying from 1% to 99%. These results show that regardless of the graph definition the number of connected components follow a similar trend as the threshold increases (Note that we exclude isolated nodes from this count of connected graph components).

The plots in Figure 4 show that the number of connected components increases up to a certain value of our similarity threshold. After a certain value of $t$, the number of connected components in the graph starts decreasing, since more and more connected components will contain only one node and will thus be excluded. The critical threshold value is different for each user similarity definition.

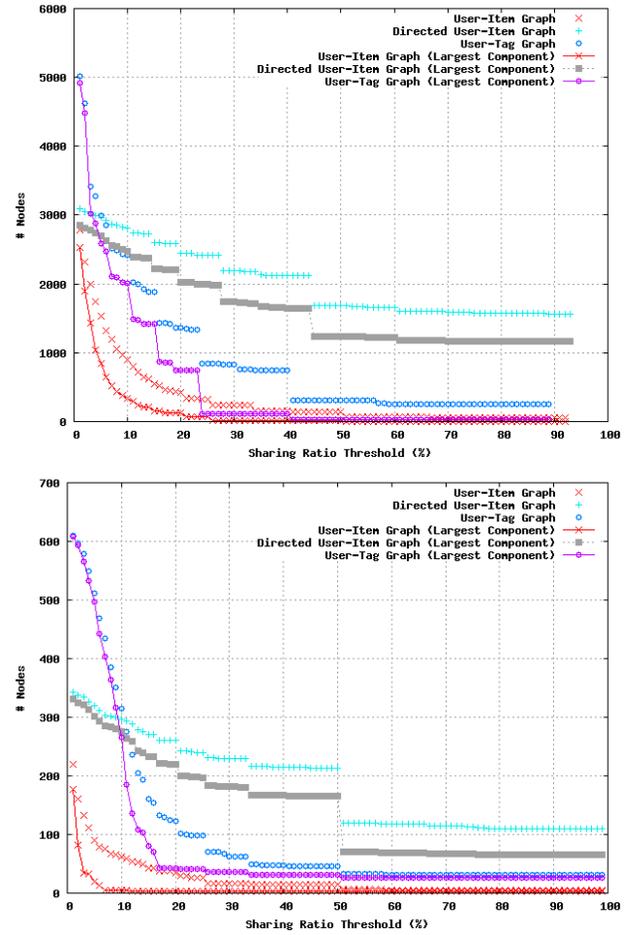

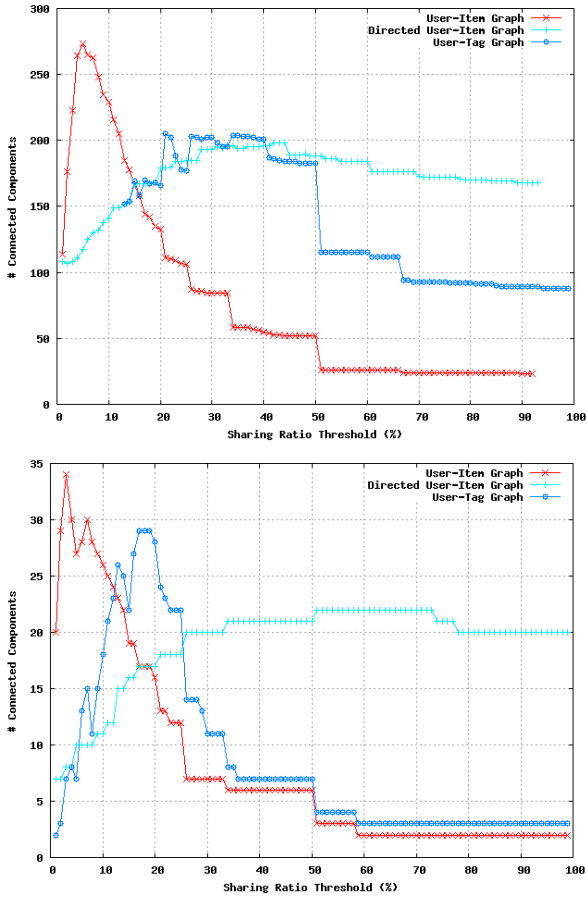

**Figure 4: Number of connected components for CiteULike (top) and Bibsonomy (bottom)**

**Figure 5: Total number of nodes in the interest sharing graph and in the largest component for CiteULike (top) and Bibsonomy (bottom)**

The initial increase in the number of connected components can be explained by the fact that, as the threshold increases, large components split to form new islands. Since these islands form naturally based on user similarity this result is encouraging since it offers the potential to cluster users according to their interests. As t continues to increase the definition of similarity becomes too strict and leads to more and more isolated nodes.

Two observations about the results in Figure 4 can be noted: first, as the threshold increases, the number of components decreases faster for the *User-Item* graph (Equation 2) than for the *Directed-User-Item* graph (Equation 4). This illustrates the effect of using an asymmetrical definition for shared interests. The idea explored here is similar to the *friendship graph,* where connections between two nodes are not reciprocal (e.g., A might consider B to be his friend, but, at the same time, B is indifferent to A) [23]. Similarly, in the directed *User-Item* graph the size of the user data set is considered leading to an asymmetrical definition of user interest (e.g., if all items of A data set are included in B's, then A will consider she has shared interests with B, while, depending of the overall size of his item set B might not consider his interests are similar enough to be connected to A's).

Second, there are more *isolated* nodes (i.e., zero-degree nodes) in the *User-Item* and *Directed-User-Item* graphs than in the *User-Tag* graph (defined by Equation 3). This indicates that users tend to have larger overlaps among their vocabularies than among their libraries. One reason for this observed interest sharing pattern may be the fact that users browse and consume items from other users' libraries without necessarily adding those items to their personal libraries in the system. An approach to verify this observation is to perform an analysis of user browsing histories to determine how often users download items from others' libraries without adding them to their personal set of items. From a system design perspective, knowing that users are more likely to share tags than data items may be useful for designing item recommendation heuristics based on vocabulary overlap.

All the similarity definitions above generally divide the original graph into one giant component, several tiny components, and a large number of isolated nodes. Figure 5 presents the total number of nodes in the components with at least two nodes and the number of nodes in the largest connected component for thresholds varying from *1% to 99%* for the three similarity measures defined above.

The results presented in this section demonstrate that using a similarity metric and the resulting interest-sharing graph it is possible to segment the user population according to manifested interest. Based on this intuition, we conjecture that it is possible to build tag/item recommendation mechanisms that exploit usage patterns, i.e., the shared interests among users. The next section offers a preliminary analysis of this hypothesis.

## 7. IMPROVING NAVIGABILITY

Chi and Mytcowicz [2] report that navigability, defined as users' ability to find relevant content, decreases as a tagging community grows. More precisely, Chi and Mytcowicz imply that the decrease in navigability is due to an increase in diversity in the set of items, users, and tags.

We have verified that the diversity of the data present in the two communities we study grows over time. Figure 6 presents the evolution of entropy, as the metric to evaluate item diversity, for the *CiteULike* data since November 2004.

As a reminder, we note that the entropy of a set of items *I*, where $|I| = N$, is defined as:

$$H(I) = -\sum_{i=1}^{N} p_i \cdot \log(p_i) \qquad (5)$$

where $p_i$ is the popularity of an item *i*.

The entropy of a set may increase in two circumstances: First, as the randomness in the item set increases (i.e., the popularity distribution gets closer to a uniform random distribution) and, second, as the set becomes larger.

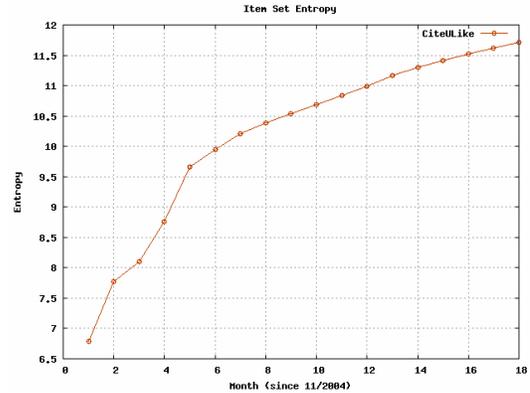

**Figure 6: Entropy growth considering items in CiteULike**

In practical terms, in a collaborative tagging community, the increase in entropy of an item set means that the user needs to filter out more items to find the one she is interested in. Similarly, high entropy makes it harder to find a tag that describes an item well. Conversely, lower entropy makes it potentially easier for a user to reach an item of interest. Thus, the question to be answered in this section is the following:

*Q3: Can the interest-sharing graph be used to reduce the entropy perceived by users when navigating through the system?*

Our two-part answer is briefly presented below and detailed in the rest of this section. First, we demonstrate that the interest-sharing graph can be used to reduce the entropy perceived by users. To this end we define a user's *neighborhood* as its set of neighbors in the sharing graph and show that this construction can be used to present users with an item set with low entropy.

Second, we offer preliminary results that suggest that this segmentation of the user population based on neighborhoods in the interest-sharing graph has a good predictive power: the items consumed by a user's neighbors predict well the future consumption pattern of that user. Thus, this offers a path to build recommendations systems based on the interest-sharing graph.

The rest of this section presents the above two-step exploration in detail. We first define a user's 'neighborhood entropy' as the entropy of the union of the item sets of that user's neighbors in the interest-sharing graph. To demonstrate that our group selection technique is effective in reducing entropy, Figure 7 compares the average neighborhood entropy in the interest-sharing graph with two types of other constructions. All results are reported with 95% confidence intervals.

Firstly, we compare for CiteULike, the average neighborhood entropy in the largest connected component of an interest sharing graph (*Average Entropy* – Figure 7), which is built by using the real tagging activity trace, to the total entropy in the system (almost double at 11.6 as presented in Figure 6). Similarly, the neighborhood entropy in the largest connected component is compared to the total entropy in the same component (*Largest*

*Component* – Figure 7). Secondly, we compare with two random graph constructions: the entropy of a random graph, which is built by selecting random neighbors from the set of users in the largest connected component (*Random Component* – Figure 7); the second random graph is built by selecting random from the entire user set (*Random Graph* - Figure 7).

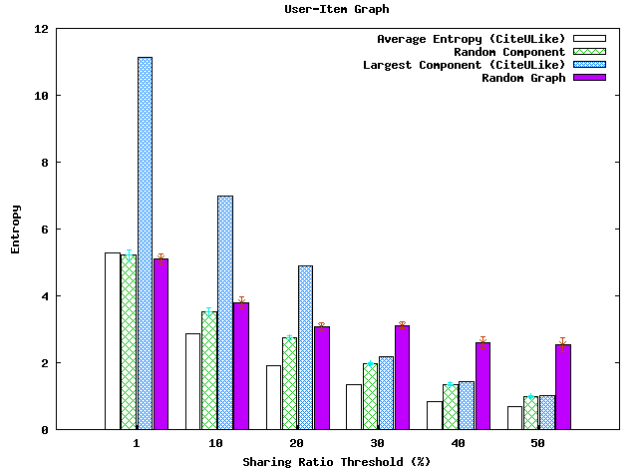

**Figure 7: The neighborhood entropy for several sharing ratio thresholds in CiteULike using *User-Item* similarity metric (Equation 2).**

We observe that the average 'neighborhood entropy' in the interest sharing graph is lower for almost all thresholds analyzed: First, the 'neighborhood entropy' in the interest sharing graph is between one half and one tenth of the entropy of the entire dataset. Additionally, for all thresholds analyzed except for the 1% threshold that does not offer enough discrimination the 'neighborhood entropy' is significantly (24% to 400%) lower than that of similar random constructions.

To support our hypothesis that the interest-sharing graph is a good basis to develop recommendation systems, we analyze how efficient the neighbor's item set in predicting future user attention over items. To this end, we evaluate the *hit ratio*: the proportion of items a user adds to her library at time *T+1* that are already in her neighbor' libraries at time *T*.

To evaluate the hit ratio, we considered the interest-sharing graph based on the User-Item similarity metric with 1% sharing ratio threshold. Preliminary results show that depending on the granularity considered (that is the length of our forecasting period: interval between *T* and *T+1*) the *hit rate* is as high as 20% for one hour granularity and decays to a low of 5% for a one-month forecast granularity. This indicates that a user's neighborhood is a possible source of information to predict near future user attention and its predictive effectiveness decreases for longer time intervals.

The findings presented on this section can be summarized as follows: First, we verify that the entropy increases as the *CiteULike* community grows overtime; Second, we verify that the interest-sharing graph can be used to reduce the entropy of item sets presented to users; and finally, we find that a user's 'neighborhood' in the interest-sharing graph is a possible source of information to predict future user actions. While the preliminary data we present here is encouraging we continue tour exploration for a complete analysis for diverse interest-sharing graphs based on multiple similarity metrics and thresholds.

## 8. CONCLUSIONS & FUTURE WORK

This work presents a characterization of two collaborative tagging systems, CiteULike and Bibsonomy, as a first step to help extract implicit information about user attention in collaborative tagging systems.

First, we analyze the distribution of tagging activity, i.e., the distribution of the volume of items, tags, and tagging actions related to each user' activity in the tagging community. We find that the activity distribution is highly heterogeneous along all these multiple axes: a few active users contribute with a large number of tag assignments and maintain a large number of items and tags, while the majority of users have a modest tagging activity.

Additionally, users with large libraries tend to have a large vocabulary of tags. While this may seem intuitive, this is not the norm across all tagging systems. In del.icio.us, for example user's library size and number of tags used are uncorrelated.

Second, we define the interest-sharing graph and investigate several definitions for interest similarity based on user activity in terms of items and vocabularies employed. Our main findings can be summarized as follows:

1. Both communities present a large population of isolated users (zero-degree nodes in the interest-sharing graph). This indicates that there are a large number of users with *unique* preferences. On the other hand, by introducing direction in the graph of shared interests, it is possible to reduce the number of isolated nodes. The final main directed connected component contains approximately twice more nodes than the undirected one.

2. The structural analysis reveals the existence of a significant number of small sub-communities of interests totally separated from each other.

3. The structure of the interest-sharing graph can be used to reduce the diversity of the items a user is exposed to. To quantify this reduction we compare the entropy of the 'neighborhood item set' with that of the item set of the entire CiteULike/Bibsonomy item set, that of the main connected component, and that of various constructions of random item sets of similar sizes.

4. Finally, we provide preliminary evidence that suggests that user's activity can be predicted by considering the union of the item sets of a node's neighbors in the interest sharing graph. We conjecture that this property can be used to build efficient, online recommendation systems for tagging communities.

This work inspires a fresh set of questions on collaborative tagging communities and contextualized attention.

A possible approach to implicitly extract user attention is to answer the following question: *What are the patterns of information propagation in collaborative tagging systems?* An answer to this question could build on previous work on information diffusion in the blogspace [7] by exploring the evolution of user attention overtime.

A second intriguing issue to explore is the following *How malicious behavior affects a tagging system and whether it be automatically detected?* Search results that are manipulated by tagging misbehavior can have an impact on usage in a collaborative tagging community [13]. Automatic detection of malicious users is paramount to the long term survival of these communities.

Finally, exploring the structure of user interest overtime can help devising models for the formation and evolution of the user similarity graphs, similarly to the study by Kumar et al. [16] on online social networks.

## ACKNOWLEDGMENTS

The authors would like to thank Richard Cameron for providing the CiteULike data set; Christoph Schmitz for providing the Bibsonomy data set; professor Lee Iverson for insightful discussions on early stages of this work, and Armin Bahramshahry, Samer Al Kiswany and Nazareno Andrade for their valuable comments. The graph analysis was executed in parallel using OurGrid (http://www.ourgrid.org).